\documentclass[10pt,conference]{IEEEtran}
\IEEEoverridecommandlockouts
\usepackage{amsmath,amssymb,amsfonts}
\usepackage{graphicx}
\usepackage{textcomp}
\usepackage[dvipsnames]{xcolor}
\usepackage{array}
\usepackage{booktabs}
\usepackage{multirow}
\usepackage{amsmath}
\usepackage{amssymb}
\usepackage{algorithm}
\usepackage[noend]{algorithmic}
\usepackage{tikz}
\usepackage{xcolor}
\usepackage{multirow}
\usepackage{paralist}
\usepackage{float}
\usepackage{makecell}
\usepackage{mathtools}
\usepackage[capitalise]{cleveref}
\usepackage{marvosym}
\usepackage{etoolbox}
\usepackage{comment}
\usepackage{listings}
\usepackage[hyphens]{url} % footnote urls, breaking lines
\makeatletter % changes the catcode of @ to 11
\newcommand{\linebreakand}{%
  \end{@IEEEauthorhalign}
  \hfill\mbox{}\par
  \mbox{}\hfill\begin{@IEEEauthorhalign}
}
\makeatother % changes the catcode of @ back to 12

\title{Simulation Environment with Customized RISC-V Instructions for Logic-in-Memory Architectures}
\author{
\IEEEauthorblockN{Jia-Hui Su, Chen-Hua Lu and Jenq Kuen Lee}
\IEEEauthorblockA{Department of Computer Science\\ 
National Tsing-Hua University, Taiwan\\
 \{chsu, kevinlu, jklee\}@pllab.cs.nthu.edu.tw}
\and
\IEEEauthorblockN{Andrea Coluccio, Fabrizio Riente and Marco Vacca}
\IEEEauthorblockA{Department of Electronics and Telecommunications Engineering\\ Politecnico di Torino, Italy\\ 
 \{andrea.coluccio, fabrizio.riente, marco.vacca\}@polito.it}
\and

\linebreakand 

\IEEEauthorblockN{Marco Ottavi\IEEEauthorrefmark{2}\IEEEauthorrefmark{4} and Kuan-Hsun Chen\IEEEauthorrefmark{4}}
\IEEEauthorblockA{\IEEEauthorrefmark{2}Department of Electronic Engineering, University of Rome Tor Vergata, Italy\\
\IEEEauthorrefmark{4}Computer Architecture for Embedded Systems, University of Twente, the Netherlands\\ 
 \{m.ottavi, k.h.chen\}@utwente.nl}

}

\begin{document}

\maketitle

\begin{abstract}
Nowadays, various memory-hungry applications like machine learning algorithms are knocking ``the memory wall''. Toward this, emerging memories featuring computational capacity are foreseen as a promising solution that performs data process inside the memory itself, so-called computation-in-memory, while eliminating the need for costly data movement. Recent research shows that utilizing the custom extension of RISC-V instruction set architecture to support computation-in-memory operations is effective. To evaluate the applicability of such methods further, this work enhances the standard GNU binary utilities to generate RISC-V executables with Logic-in-Memory (LiM) operations and develop a new gem5 simulation environment, which simulates the entire system (CPU, peripherals, etc.) in a cycle-accurate manner together with a user-defined LiM module integrated into the system. This work provides a modular testbed for the research community to evaluate potential LiM solutions and co-designs between hardware and software.

%with a standard memory interface, by which different logic-in-memory architectures can be hooked into the microprocessor directly.

\thispagestyle{plain}
\end{abstract}
\section{Introduction}
\label{sec:introduction}
%\kuan{Page 1.5}

The performance gap between memories and computing units, known as ``the memory wall'' has been a long-lasting issue in computing systems~\cite{10.1145/977091.977115}. Along with the rise of application domains such as machine learning and the internet of things, this issue becomes 
exceedingly pressing in the conventional von Neumann computing paradigm. In addition to the ever-worsening performance gap, the data movement between these components consumes more than 60\% of the total system energy~\cite{10.1145/3173162.3173177}. To overcome this issue, several Computation-in-Memory (CiM) techniques have been proposed, and research from different computing layers is actively involved~\cite{9473976, 9643588}. Specifically, Logic-in-Memory (LiM) is one promising CiM solution~\cite{jlpea10010007, electronics11192990}, where the computation can be performed by additional logic inside the memory array\footnote{The literature offers a wide range of LiM definitions~\cite{9473976, mi10060368}, and we follow the typology used in~\cite{mi10060368} and~\cite{electronics11192990}. Nevertheless, this work's insight shall generally apply to memory architectures with specific computation capacities.}.

%\kuan{Motivate LIM, due to memory-wall, low-power, emerging memories}

%As an open standard, RISC-V instruction set architecture (ISA) offers the flexibility for researcher and designers 
\begin{figure}[t]
    \centering
    \includegraphics[width=0.95\textwidth/2]{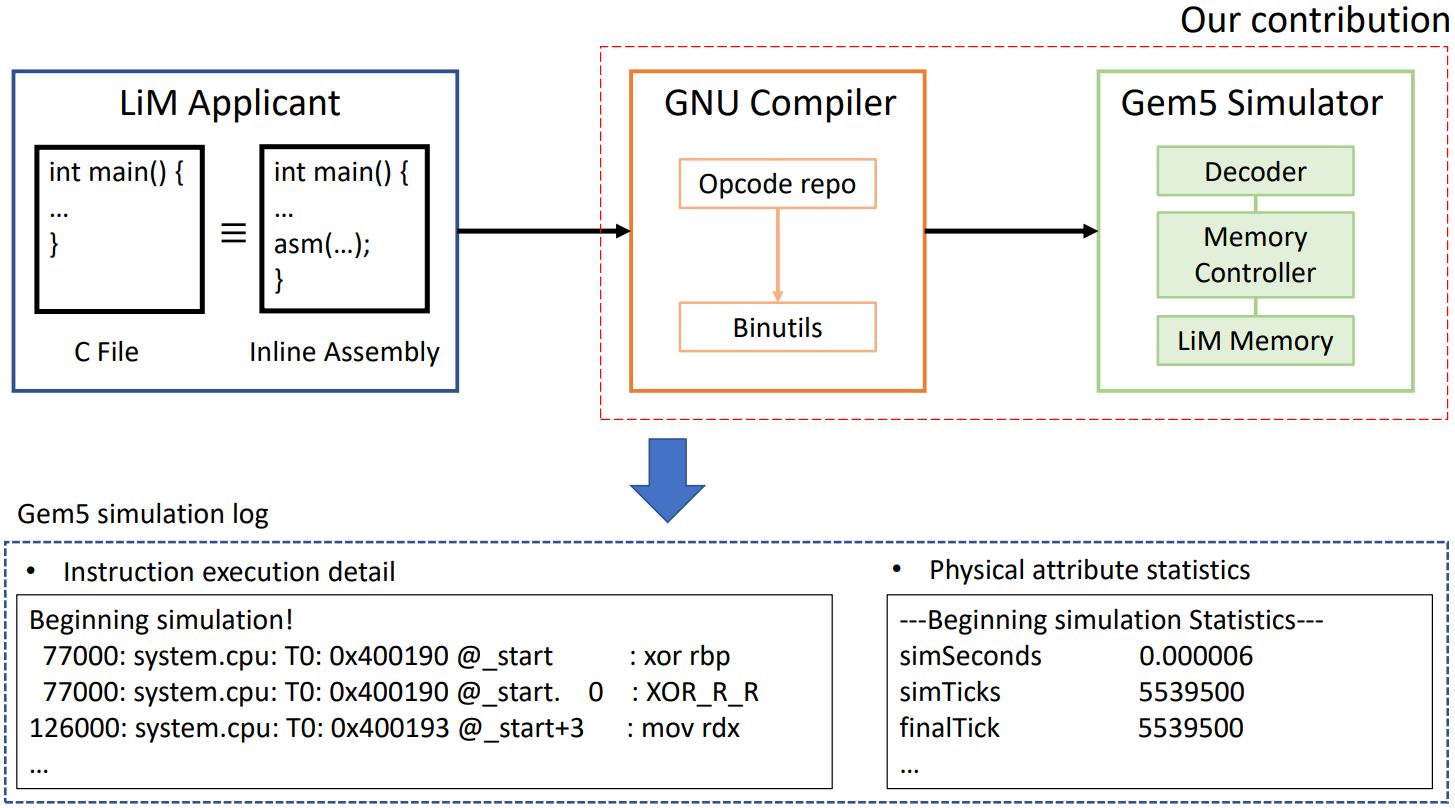}
    \caption{Overview of the evaluation environment developed in this work. Our contributions are within the scope of the dashed block. The outputs of the environment include simulation logs and instruction execution logs.
    %\su{Do we need to talk about the log in the caption?}
    %\kuan{Should we put the two log blocks in the evaluation part? And make those text bigger}
    }
    \label{fig:contribute}
    \vspace{-0.5cm}
\end{figure}
To utilize such CiM architectures, the CPU must be able to coordinate the corresponding operations and let the data processing perform by the memory itself. A few recent studies have shown that the custom encoding space of RISC-V Instruction Set Architectures (ISA) serves this purpose very well, in which additional instructions can be introduced and mapped to various functionalities provided by CiM architectures. For example, Lin et al.~\cite{10.1145/3458744.3473351} propose specific instructions with the support of Tensor Virtual Machine (TVM) to aid Processing-in-Memory architectures for accelerating binarized neural network~\cite{DBLP:journals/corr/abs-1802-04799}. Considering the standard RISC-V core architecture~\cite{ri5cy}, RISC-Vlim provides a general solution to enable the communication between the microprocessor and the LiM array without altering the bus interface~\cite{electronics11192990}. With such customized instructions, the coordination of the computing process can be automated, and more optimizations are potentially allowed.
%\kuan{Why RISC-V for LIM? What have been shown? The one with PIM on Gem5 (as the closet work)?}

However, computing systems with the CiM capacities are still rarely available at the market. With commodity memories like DRAM, indeed some prominent examples have been emerging, e.g., Samsung's FIMDRAM~\cite{10.1109/ISCA52012.2021.00013}, UPMEM architecture~\cite{9651614}, SK Hynix's AiM~\cite{9731711}. Most researchers so far still rely on simulations to evaluate their approaches, based on detailed models at different layers. Under this context, the gem5 simulator~\cite{10.1145/2024716.2024718} is widely used nowadays and serves several research directions, especially for emerging memories and HW/SW co-designs~\cite{10.1145/3483839, 8344612, 9045418}. It is a cycle accurate full-system simulator, which provides the functional simulation of the underlying hardware. Many simulation tools are also possible to be integrated as modules or plugins with the gem5 simulator, such as NVMain2.0~\cite{7038174} for non-volatile memories and RTSim~\cite{8642352} specifically for racetrack memories. \newline

% the gem5 simulator~\cite{10.1145/2024716.2024718}, which is a cycle accurate full-system simulator, is widely used nowadays
% \kuan{What could be advantages if we have Gem5  simulation? Page 4}
% Why we simulated Logic in Memory (LiM) instructions with Gem5? 
% % Full system simulation
% First, Gem5 supports the full system simulation. When a program is executed, the instruction count profiling of the circuit is important. However, the issues with the operating system are also needed to be considered. Take consideration of the memory issues, if a program is executed without page fault, and interrupt, the measurement result has a large difference from the actual situation.
% % Efficiency
% Second, Gem5 is more efficient than some low-level simulators. Gem5 takes the hardware components, e.q. CPU, memory, and cache, as objects. The communication between these components is described in an objective-oriented way, instead of a circuit simulation.
% % Low-level support
% Third, Gem5 gives comprehensive support with low-level functions. While debugging, Gem5 allowed users to print necessary information, including the register data changed step by step.
% \kuan{Draw a diagram based on the flow of Page 7}

%Figure~\ref{fig:lim_flow} described our LiM instruction simulation flow. First, we have a C code suitable for LiM application. We rewriting this program with inline assembly function which embedded the customized LiM assembly code into the program. Then, we compile this program with modified GCC and executed the process with Gem5.

\noindent\textbf{Our Contributions:} In this work, we develop a new simulation environment, where the customized RISC-V instructions defined in~\cite{electronics11192990} can be integrated into a C program via the inline assembly functions, and the developed gem5 simulator is able to run the generated executable and report the execution details, such as execution time in cycles and executed instructions logs. Figure~\ref{fig:contribute} shows the overview of this work. 
In a nutshell, the contributions of this work are:
\begin{itemize}
    \item Introducing the customized RISC-V instructions, defined by~\cite{electronics11192990}, into the standard GNU binary utilities, by which the inline assembly functions in the C program for LiM operations can be compiled (See Section~\ref{sec:binutils}). 
    %{Modify RISC-V GNU toolchain with customized instructions -- so that a source code with inline assembly functions}
    \item The gem5 simulation setup with an enhanced decoder for the customized RISC-V instructions, which simulates the LiM operations with a user-defined LiM module on a functional level (See Section~\ref{sec:gem5}).
    %{Introduce LiM customized instructions to Gem5 simulator}
    \item Evaluation of the developed simulation environment with real-world algorithms, in comparison to the simulation results derived from RISC-Vlim~\cite{electronics11192990} (See Section~\ref{sec:validation}).
    %{Validation between POLITO repos and the developed Gem5, Detailed instruction execution logs, Execution time in cycle}
\end{itemize}

\begin{figure}[t]
    \centering
    \includegraphics[width=0.95\textwidth/2]{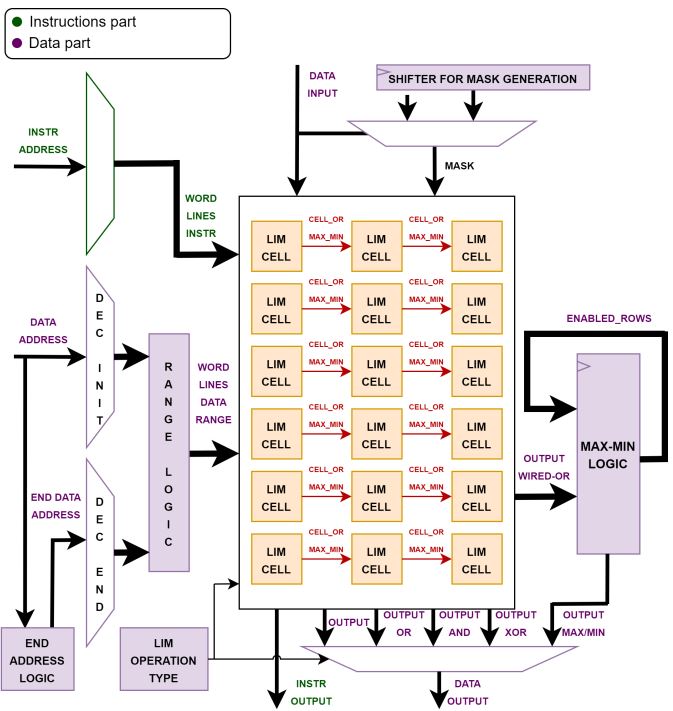}
    \caption{LiM memory architecture, adapted from~\cite{electronics11192990}}    \label{fig:italyMem}
\end{figure}

\begin{figure}[t]
    \centering
    \includegraphics[width=0.6\textwidth/2]{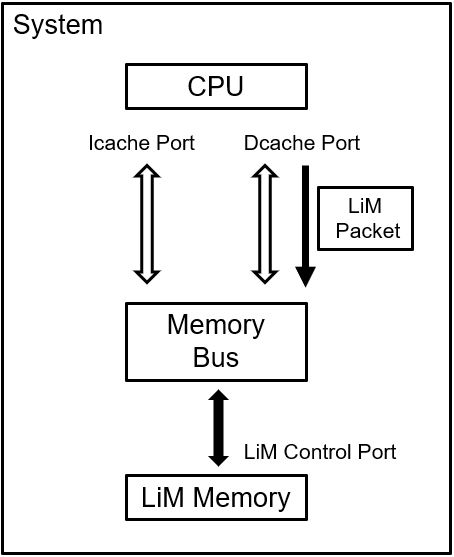}
    \caption{The considered system architecture, simulated in the developed gem5 simulation environment. As a proof of concept, we disable the cache hierarchy in this work and focus on the CPU and LiM modules.}
    \label{fig:arch}
\end{figure}

The enhanced toolchains and the developed environment  in this work will be publicly available once the paper is accepted.
%\kuan{Source code release}

\section{System Architecture and RISC-V Instructions}
%\kuan{Page 0.5 --  2 paragraphs + 2 figures}
In this section, we introduce the system architecture and the customized RISC-V instructions considered in this work. We follow the insight of Coluccio et al.~\cite{electronics11192990} to consider a standard memory interface, by which different LiM architectures are compatible, regardless of the underlying memory technologies. 

\subsection{System Architecture}
%\kuan{System architecture, Page 12}
% Paragraph described Italy Memory implement  
Figure~\ref{fig:italyMem} shows the considered LiM memory design~\cite{electronics11192990}. Noted that the CPU used in~\cite{electronics11192990} ri5cy access instruction and data from the same memory, so the instruction flow is also considered in the memory design. Each memory cell is considered a LiM cell, while access data, bit-wise operations like \textbf{AND}, \textbf{OR}, and \textbf{XOR} will be operated with the mask data, which is from the register. All the LiM cells are connected with the MAX-MIN Logic, which determines the MAX or MIN value over the cells. In this simulation, we realize the functionality of range logic with its peripherals and LiM cells. MAX-MIN logic is considered as the future work.

Figure~\ref{fig:arch} illustrates the system architecture we considered in this work. Based on the default design of gem5, the communication over the hardware components as modules in the simulated system is implemented via packets and ports. The CPU has the instruction cache port and the data cache port connected with the memory bus, and the memory bus received data from CPU and control the LiM memory architecture. The corresponding portions of the LiM memory can then be activated via the information contained in the packet, such as opcode and offset of the memory address (introduced below).
Any further interface modification, such as the memory interface, can also be easily introduced.
Although gem5 has memory subsystems such as cache hierarchy, they are disabled in this work to avoid involving side effects, which are considered out of the scope.

% \kuan{We haven't introduced the instructions yet from the reader's flow!!}
% Once \textbf{STORE\_ACTIVE\_LOGIC} is executed, CPU send a packet with the data in \textit{BASE\_REG}, \textit{RANGE\_REG} and \textit{MEM\_OP}. 

\subsection{Customized RISC-V Instructions for LiM Operations}

\begin{figure*}[t]
    \centering
    \includegraphics[width=0.6\textwidth]{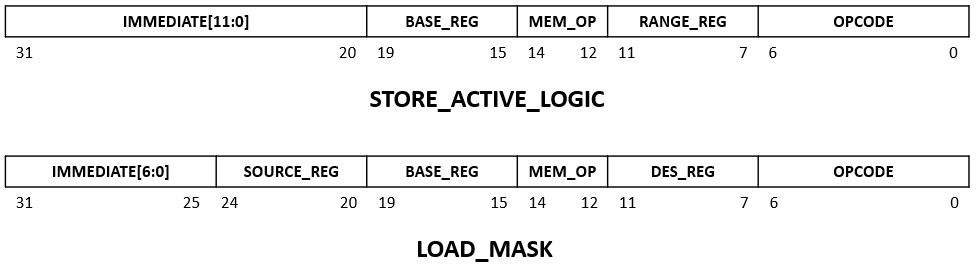}
    \caption{RISC-V customized instructions to use the LiM memory cells, adapted from~\cite{electronics11192990}.}
    \label{fig:insts}
\end{figure*}

For the completeness, we recap the customized RISC-V instructions introduced in~\cite{electronics11192990}, which support the LiM solution proposed in~\cite{8617879}.
Since the LiM memory could also be
utilized as a classic memory, two customized instructions are introduced for adjusting the status of the LiM memory cells, namely \textbf{STORE\_ACTIVE\_LOGIC} and \textbf{LOAD\_MASK}, as shown in Figure~\ref{fig:insts}. The original \textbf{STORE} instruction could also be involved in the operations of LiM process. If the memory region is activated for LiM operations, a normal store instruction will be interpreted as a logic store instruction.  

There are three fields for \textbf{STORE\_ACTIVE\_LOGIC}: \textit{BASE\_REG}, \textit{RANGE\_REG} and  \textit{MEM\_OP}. The \textit{BASE\_REG} field determines the base memory address, and the range size (i.e., \textit{RANGE\_REG}) of cells are active. The \textit{MEM\_OP} field gives the active operation type: \textit{NONE}, \textit{AND}, \textit{OR}, \textit{XOR}, \textit{NAND}, \textit{NOR} and \textit{XNOR}. For \textbf{LOAD\_MASK}, data from the \textit{BASE\_REG} field will be masked by the \textit{MEM\_OP} with the input mask in the \textit{SOURCE\_REG} field first. Afterward, the result is stored at the destination register \textit{DEST\_REG}. Note that \textbf{LOAD\_MASK} must always be placed after the activation of the LiM operation, which is handled by compilation.

\begin{figure}[t]
    \centering
    \includegraphics[width=\textwidth/2]{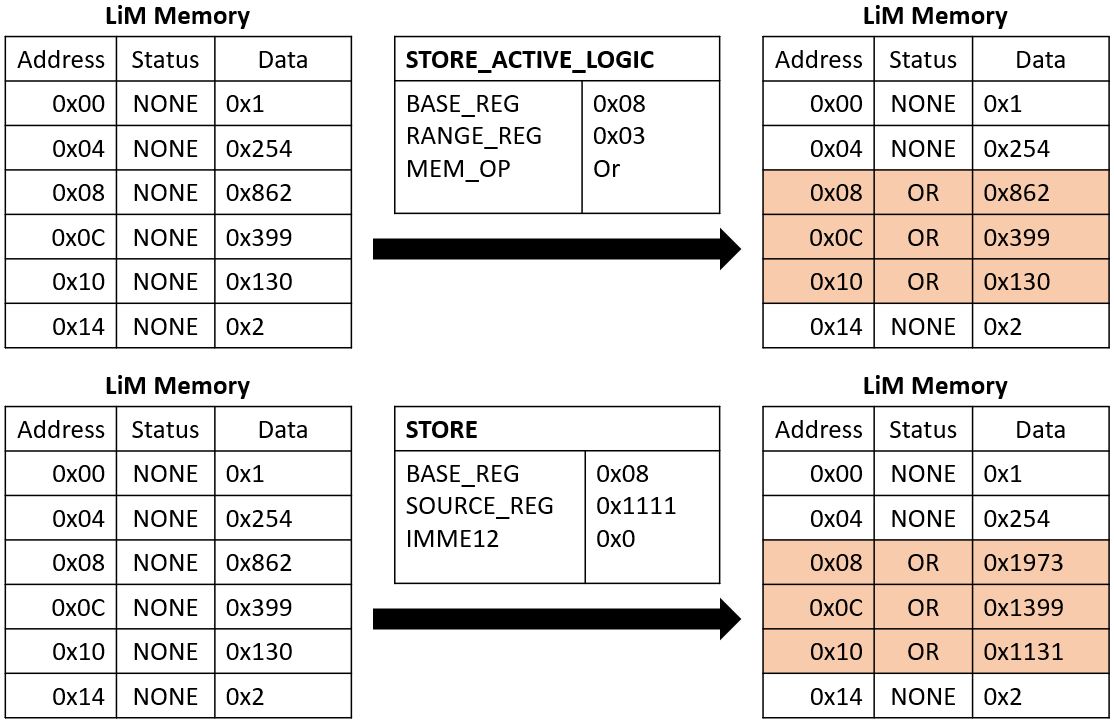}
    \caption{Running example for \textbf{STORE\_ACTIVE\_LOGIC}}
    \label{fig:sw_ex}
\end{figure}

% \kuan{recap the LIM paper}

% \kuan{Page 13 and 14 can be used as examples to showcase the behavior of each introduced instruction}

Here we give a running example to demonstrate the system behavior with the instruction \textbf{STORE\_ACTIVE\_LOGIC} with \textbf{STORE}. Suppose that every LiM memory cell has an initial status, \textbf{NONE}, meaning that this cell is inactivated for any LiM operation yet and data is merely loaded into the memory. 
As shown in Figure~\ref{fig:sw_ex}, \textbf{STORE\_ACTIVE\_LOGIC} executed with \textit{BASE\_REG} equal to \textit{0x08}, \textit{RANGE\_REG} equal to \textit{0x03}, and \textit{MEM\_OP} is \textit{OR}. The LiM memory switches the cell status to \textit{OR} for 3 cells starting from the address \textit{0x08}. The \textbf{STORE\_ACTIVE\_LOGIC} instruction must cooperate with a 
\textbf{STORE} instruction, which is treated as a logic store operation here.
Note that the \textbf{BASE\_REG} field must have the same value and the value of \textit{IMME12} field must be zero. Therefore, the data from the \textit{SOURCE\_REG} then are operated with \textit{MEM\_OP} and written back to the same memory cell. 
%\kuan{This example doesn't use LOAD\_MASK?} %The data from the \textit{SOURCE\_REG} do the \textit{MEM\_OP} with the original data in the cell and write back to the cell.

\subsection{Enhancement of GNU Binary Utilities}
\label{sec:binutils}

%\kuan{GCC binutil to define opcodes with the additional instructions}
% \kuan{Page 0.5}

Here we present how we generate the executable binary with the customized instructions via the enhancement of the GNU binary utilities. We consider an agile development flow with the usage of inline assembly functions to enable the LiM operations in a given program.

In a typical end-to-end compilation flow, the source code is first translated into the assembly code and converted into the object code by the assembler. Afterward, the linker binds and links runtime information to generate the executable binary. 
To serve the objective of this work, the enhancement of the assembler is sufficient without revamping all the steps above. 

%\kuan{Need to explain to which extent we are similar as Section 6 of the LiM paper.}

% 

% \kuan{First explain why we use binutil to work. The reason that we can avoid any effort for runtime (linking, binding) that we use inline assembly functions.}

% In the end-to-end compilation flow from the source code to the executable binary, the source code is translated into the assembly code first, then the assembly code is converted into the object code by the assembler. Finally, the linker binds and links runtime information to make the executable program. 

%When a new instruction is introduced, a more agile development flow by inline assembly function is helpful. We described our LiM application by C program with inline assembly functions. With the GNU compilation flow, only the assembler needed to be modified with new instructions.

\begin{figure}[t]
    \centering
    \includegraphics[width=\textwidth/2]{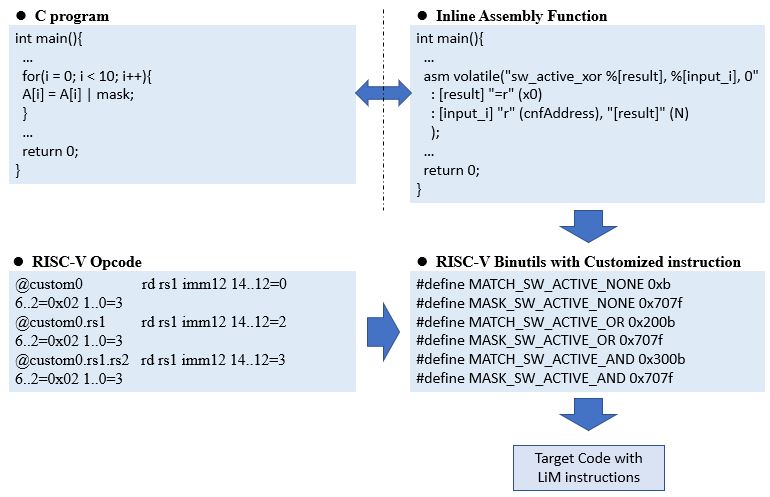}
    \caption{Overview from a given C program to the target code}
    \label{fig:lim_flow}
\end{figure}

% \kuan{To make it work, what needs to be done}
%\kuan{I will continue here.}
Figure~\ref{fig:lim_flow} illustrates the compilation flow we adopted: Given an application source code, the LIM operations are enabled by users through inline assembly functions. To process the additional customized instructions, the new opcodes have to be registered in the assembler. The RISC-V GNU binary utilities provide the repository of RISC-V Opcode to reserve the opcodes for customized instructions in the repository:.
Since there is no automatic detection for collisions, a potential pitfall here is that the introduced opcodes might overlap with the existing opcodes. 
With the auxiliary of the RISC-V opcode, we can introduce new instructions into the GNU binary utilities with unique opcodes. 
%Note that it is important to avoid overlapping with the normal RISC-V instructions. A potential pitfall here is that 

%We have an application suitable for the architecture, and we described that by inline assembly functions. By registering the new instruction format, the assembler recognizes the new instruction and generates the corresponding target code with LiM instructions described in Section~\ref{sec:introduction}. To register the new instruction format, finding a new opcode not overlap with the current instruction is important, and the RISC-V GNU toolchain provides the RISC-V Opcode \su{cite RISC-V opcode?} repository reserve the opcode for customized instructions. By the auxiliary of the RISC-V Opcode, we introduce new instructions into the GNU Binutils with unique opcode, and the assembler compiled our C code with inline assembly functions into the executable binary run on Gem5. 

\section{Gem5 Simulation Environment}
\label{sec:gem5}

In this section, we present the design of our gem5 simulation environment, where the RISC-V decoder is extended to process additional instructions introduced previously. Figure~\ref{fig:lim_gem5} shows the overview, considering one LiM architecture in the simulated system suggested by~\cite{electronics11192990}.
%\kuan{Page 1}

%The compilation flow for customized instructions is described in Section~\ref{sec:binutils}, Gem5 simulator for LiM instructions is introduced in this section.

\subsection{RISC-V Decoder for LiM Instructions}
\label{sec:decoder}

%I will start from this

\begin{figure*}[t]
    \centering
    \includegraphics[width=0.7\textwidth]{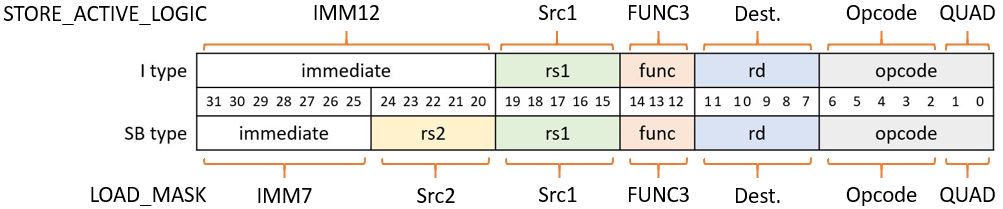}
    \caption{RISC-V instruction type and corresponding Gem5 decode format}
    \label{fig:decoder}
\end{figure*}

To use the LiM memory architecture, at first two new customized instructions, \textbf{STORE\_ACTIVE\_LOGIC} and \textbf{LOAD\_MASK} are additionally introduced into the RISC-V decoder of gem5. Under 32-bit RISC-V ISA, \textbf{STORE\_ACTIVE\_LOGIC} belongs to I-type, and \textbf{LOAD\_MASK} is the SB-type~\cite{isaParser}.  
Figure~\ref{fig:decoder} shows the instruction format of RISC-V I type and SB type. 

The ISA Parser (so-called decoder) adopts a nested structure to decode instructions, based on the code segments (i.e., the brackets in Figure~\ref{fig:decoder}). %Figure~\ref{fig:code} gives an example of how the instruction format of \textbf{STORE\_ACTIVE\_LOGIC} is registered.  
The format is defined from left to right and match the \textbf{QUADRANT}, \textbf{OPCODE}, and \textbf{FUNC3}. For all the instructions, the memory and the registers exchange data to each other. For \textbf{STORE\_ACTIVE\_LOGIC}, the activation size of memory stored in the \textit{RANGE\_REG} is sent to the memory via the packet, i.e., \textit{Mem\_ub} is assigned with \textit{Rd\_ub}. For the simplicity of presentation, the other detailed handling such as the memory address in \textbf{RANGE\_REG} and the \textit{MEM\_OP} are not introduced here.

\subsection{Design of LiM Module}
\begin{figure*}[t]
    \centering
    \includegraphics[width=0.7\textwidth]{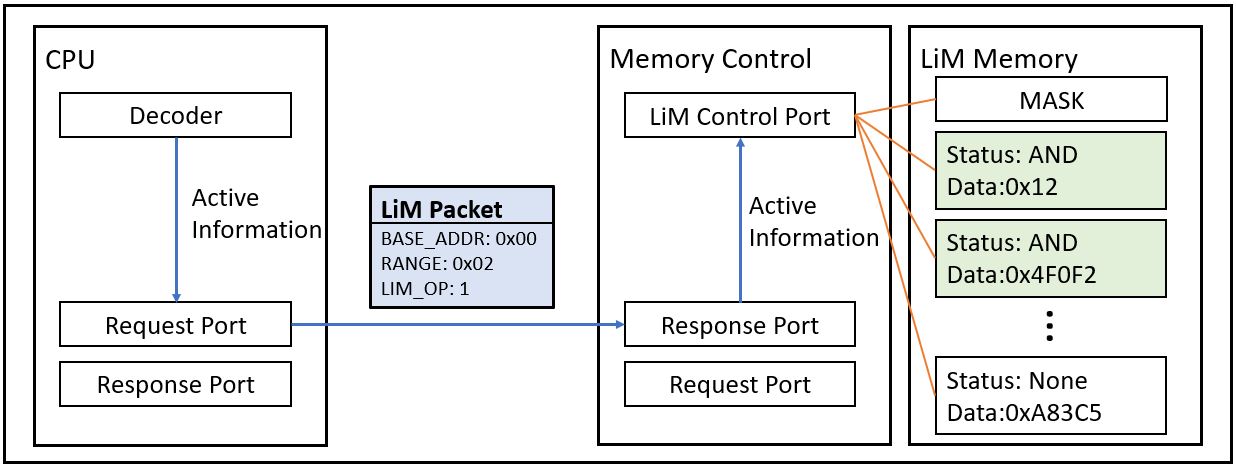}
    \caption{The developed gem5 simulation environment, integrated with the considered LiM memory architecture}
    \label{fig:lim_gem5}
\end{figure*}
Along with the convention of gem5, we introduce a module to describe the functionalities of the LiM memory architecture into the environment, as shown in Figure~\ref{fig:arch}, where the CPU communicates with the memory controller through the packets. 
Based on the information contained in the packets, the memory controller modifies the state of the memory cells accordingly and updates the data stored in the memory. 

%\su{Describe Gem5 simulate}
Figure~\ref{fig:lim_gem5} illustrates the gem5 simulator, integrated with one LiM memory architecture. By the design of gem5, every hardware component, such as CPU, is considered as an object, and the objects communicate with each other via packets and signals. After decoding, all the information is packed into the LiM packet, and the CPU module sends this packet to the memory controller by the response port. Afterward, the memory controller alters the state of the LiM memory. Each cell inside the memory holds the current states of LiM memory as \textit{NONE}, \textit{AND}, \textit{OR}, \textit{XOR}, \textit{NAND}, \textit{NOR}, or \textit{XNOR}. After the simulation, instruction count, simulation time, and detailed instruction log is obtained by the gem5.

\section{Evaluation and Discussion}
\label{sec:validation}

In this section, we present the required runtime for simulating the execution of benchmarks with the introduced customized instructions, in comparison to the RTL simulation environment provided by~\cite{electronics11192990}. Please note that, 
%While the speedup over the RTL simulations in
%Please note that, 
%the purposes of two simulation environments are essentially different.  
the simulations conducted in~\cite{electronics11192990} are based on Vsim (as known as ModelSim), where a hardware description language is utilized and waveforms for signals are generated for circuit validation. Once the designed architecture is proven to work well with the customized instructions, the gem5 simulation environment developed in this work can be utilized to efficiently evaluate the effectiveness of customized instructions and LiM architecture. The resulted speedup is thus expected and helpful.
%Therefore, this work is orthogonal to~\cite{electronics11192990}.

%To run the process on Vsim and Gem5, the inline assembly in C is different. 

\begin{figure}[t]
    \centering
    \includegraphics[width=\textwidth/2]{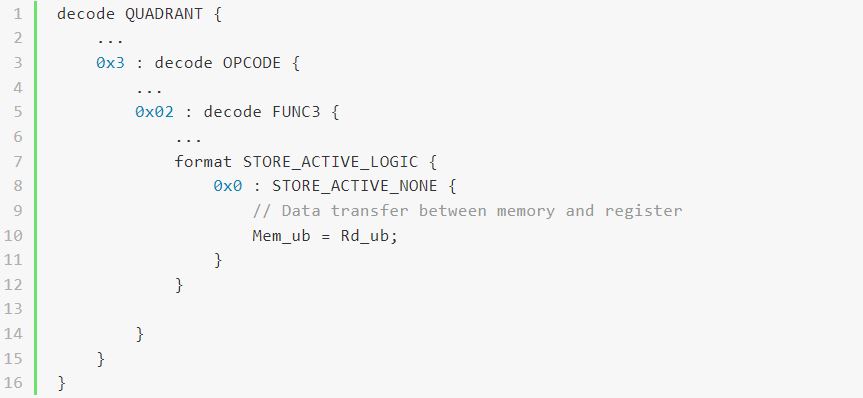}
    \caption{Gem5 decoder example for \textbf{STORE\_ACTIVE\_LOGIC}}
    \label{fig:code}
\end{figure}
\subsection{Experimental Setup}

Table~\ref{tab:env} shows the details of experiment environment. We simulated a system with a LiM memory architecture in Gem5 v22.0.0.2. The customized instructions were introduced by the enhanced GNU Binutils 11.1.0 (see Section~\ref{sec:binutils}).
We compare the performance of the simulation environments with the same benchmarks adopted in~\cite{electronics11192990}, i.e., The source code with inline assemblies were provided by the repository\footnote{\url{https://github.com/vlsi-nanocomputing/risc-v-lim-architecture}}, and we used the enhanced GNU Binutils to generate the target code in \texttt{elf} with the LiM operations as the input of the gem5 simulations.

% \kuan{We run on what machine, size of memory, which CPU, to measure the runtime. What are the benchmarks we considered?}

% \subsection{Memory Space in Vsim-and gem5-based Simulations}

Since the Vsim-based evaluation conducted in~\cite{electronics11192990} merely simulates the hardware behavior, only the physical memory space is considered, and thus an actual physical address is given into the tested application, shown as Listing~\ref{lst:vsim}. 
However, our gem5-based simulation environment works with a virtual memory space towards the need of operating systems, where we adopt the \textit{malloc} function to settle the virtual memory space, as shown in Listing~\ref{lst:gem5}. 

\begin{lstlisting}[language=c++, caption={Address definition example for vsim}, basicstyle=\footnotesize, captionpos=b, frame=single, label=lst:vsim]
int main(int argc, char* argv[]) {
    /* input variables declaration */
    int (*states)[4][4] = 0x30000;
    ...
    return EXIT_SUCCESS;
}
\end{lstlisting}

\begin{lstlisting}[language=c++, caption={Address definition example for Gem5}, basicstyle=\footnotesize, captionpos=b, frame=single, label=lst:gem5]
int main(int argc, char* argv[]) {
    /* Other variables */
    int i;
    /* input variables declaration */
    int row = 4, col = 4;    
    int* states[row];
    for (i = 0; i < row; i++)
        states[i] = (int*)malloc(col * sizeof(int));
    ...
    return EXIT_SUCCESS;
}

\end{lstlisting}
\begin{table}[t]
    \centering
\begin{center}
\begin{tabular}{ p{\textwidth/6} p{\textwidth/4} }
\hline
\multicolumn{2}{c}{\textbf{Experiment environment}} \\
 \hline 
 Architecture&              x86\_64 \\
 CPU(s)&                        24 \\
 Model name&             12th Gen Intel(R) Core(TM) i9-12900K \\
 CPU max MHz&           6700 \\
 CPU min MHz&            800 \\
 % L1d cache&                   640 KiB \\
 % L1i cache&                   768 KiB (16 instances) \\
 % L2 cache&                     14 MiB (10 instances) \\
 % L3 cache&                     30 MiB (1 instance) \\
 Main Memory Size&                  128 GiB \\
 Operating System&     Ubuntu 22.04.1 LTS \\
 \hline
\end{tabular}
\end{center}
\caption{Experiment environment}
\label{tab:env}
\end{table}
\subsection{Experimental Results}
Table~\ref{tab:time} shows the required time of two simulation environments for running different benchmarks on average. For each simulation environment, we ran $20$ times per benchmark. We can see that for running software applications, the gem5-based environment is much efficient. For \texttt{xnor\_net.c}, due to the limited time, we cannot finish all the experiments.
%Table~\ref{tab:cc} reports the clock cycles of each benchmark. Since the simulations were conducted in the syscall mode (SE), where the gem5 simulator emulates the process of Linux system call to run the program, the clock cycles in statistics in the gem5 is larger than Vsim, which simulated the bare metal process. 

%\kuan{We need to explain why the cc of gem5 is much bigger. If Vsim runs the same apps with LiM instructions we run, the number should be the same of similar scale.}

%We executed the benchmark execution with the system call emulation on Gem5, under this mode, Gem5 simulates the process of the operating systems make a system call to run the program. 
% The overhead of memory allocation and memory control, such as heap and stack operations, also takes into consideration. Thus, the clock cycles in statistics in Gem5 is larger than Vsim, which simulated the bare metal process. 

Overall, as long as the introduced customized instructions have been validated by, for example, the Vsim environment in~\cite{electronics11192990}, the gem5-based environment is more suitable for massive testing for capturing the numbers of memory accesses and the performance improvement.

%\kuan{Page 1}
%\kuan{Briefly talk about what kind of evaluation has been done in POLITO's repository.}
% \kuan{We present the required time for analysis}
% \kuan{Discuss what we get with the current version, and what could be further done -- full system, by which what else can be evaluated and so on. Operating systems can also be involved.}

\section{Related Work}
Several CiM solutions have been proposed~\cite{9473976, 9643588}, especially in recent years due to the emerging non-volatile memories~\cite{9712536, 9218590, 8980300}. Using customized RISC-V instructions to support various architectures such as accelerators has also been studied widely~\cite{10.1145/3458744.3473351, electronics11192990}. 
RISC-V is conceived to be expandable so that unused opcodes can be introduced to accommodate custom extensions; some extensions become standard, while others can remain a specific feature of vendors. This characteristic allows RISC-V processors to have ad-hoc instructions while keeping compatibility with the global RISC-V ecosystem. This approach has been, for example, leveraged by the PULP processors \cite{7864441} \cite{8714897} \cite{9068465}. 

%\kuan{Marco O., please add 2 papers such as using the RISC-V custom extension to support on-board architectures from ESA?}. 
While several instruction set simulators for RISC-V have been developed to speed up the functional verification, most of them can hardly be extended to support system-level use cases. Different full system simulators have been proposed to fill this gap, such as SoCRocket~\cite{2013ESASP.720E} and RISC-V VP~\cite{HERDT2020101756}, just to name a few. As one popular full system simulator, gem5 has been attractive to computing system researchers~\cite{10.1145/2024716.2024718, 9045418, 9984978}. This work serves as the first step to support the customized RISC-V instructions in the gem5 fashion for LiM memory architectures, by which more LiM solutions, HW/SW co-designs and software automation can be evaluated. 
% \kuan{0.25 page - 1 paragraph}
% \kuan{Gem5-based PIM framework, LIM}
\section{Conclusion}
% \kuan{0.25 page - 1 paragraph}
% \kuan{Fairly describe what we want to achieve, and what have been achieved so far. Page 16 for future work.}
Towards the pressing issue of "the memory wall", various CiM solutions have been proposed in the literature. Recent research shows that utilizing the custom extension of RISC-V instruction set architecture (ISA) to support the coordination of CPU and CiM, especially for LiM architectures, is effective. In this work, we develop a new gem5 simulation environment to provide a cycle-accurate simulation to support such research. The results show that the gem5-based simulation can be used for massively testing the impacts of introduced instructions, and its flexibility provides foreseen potentials for testing CiM/LiM solutions. In the future, we plan to maintain this tool set and include, for example, more LiM architectures and customized instructions like reduction algorithms.

\begin{table}[t]
    \centering
\begin{center}
\begin{tabular}{ p{\textwidth/8} p{\textwidth/8} p{\textwidth/8} }
\hline
\multicolumn{3}{c}{\textbf{Simulation time}} \\
 \hline 
 Benchmark          & Gem5      & Vsim \\ 
 \hline
 aes128\_arkey.c    & 0.0149s   & 53s \\ 
 bitmap\_search.c   & 0.0148s   & 57s \\
 bitwise.c          & 0.013s    & 36s \\
 max\_min.c         & 0.0232s   & 116s \\
 xnor\_net.c        & 0.0886s   & -\\
 \hline
\end{tabular}
\end{center}
\caption{Simulation time comparison}
\label{tab:time}
\end{table}

% \begin{table}[t]
%     \centering
% \begin{center}
% \begin{tabular}{ p{\textwidth/8} p{\textwidth/8} p{\textwidth/8} }
% \hline
% \multicolumn{3}{c}{\textbf{Clock cycles [cc]} }\\
%  \hline 
%  Benchmark          & Gem5    & Vsim \\ 
%  \hline
%  aes128\_arkey.c    & 16705   & 529 \\ 
%  bitmap\_search.c   & 16882   & 454 \\
%  bitwise.c          & 8866    & 332 \\
%  max\_min.c         & 95206   & 381 \\
%  \hline
% \end{tabular}
% \end{center}
% \caption{Clock cycles comparison}
% \label{tab:cc}
% \end{table}

% \section*{Acknowledgement}
% \kuan{If we need something to top up the space to 6 pages, here is the place.}

\bibliographystyle{abbrv}
\bibliography{main}{}
\end{document}